# Extending the thermal near field through compensation in hyperbolic waveguides

Sean McSherry and Andrej Lenert

*Department of Chemical Engineering, University of Michigan, Ann Arbor, MI, 48109*

E-mail: mcsherry@umich.edu; alenert@umich.edu

## ABSTRACT

A promising method to leverage near-field power densities without the use of nanoscale vacuum gaps is through hyperbolic metamaterial (HMM) waveguides. When placed between a hot and cold reservoir, an ideal HMM can transmit surface waves across several microns, enabling an extension of near-field enhancements. However, when accounting for transmission loss due to realistic levels of absorption within the waveguide, previous studies have shown that the enhancements are significantly curtailed at wide separations. In our study, we investigate the role of internal sources within realistic non-isothermal HMMs. We demonstrate that, in some cases, the emission from the HMM accounts for over 90% of the total radiative heat transfer to the receiver, and that these additional sources can largely compensate for optical losses associated with decreased transmission from the emitter to the receiver. Lastly, we investigate the spectral transport in a realistic 3-body system that has mismatching optical properties between the boundaries (emitter and receiver) and the waveguide (HMM). Our model shows that the near-field thermal transport remains spectrally selective to the boundaries, even as major radiative contributions come from the waveguide. This work may enable the design of non-isothermal emitter-waveguide-receiver systems that transmit near-field power levels over wider separations.

## I. INTRODUCTION

Radiative transport between materials separated by distances less than the thermal wavelength (i.e., near-field) exceeds the far-field blackbody limit due to coupled evanescent waves [1]. The increased power density [2,3] and frequency tunable [4] nature of near-field transport may benefit technologies such as thermo-photovoltaic/photonic/radiative devices [5–8] and infrared sensing [9]. Proof-of-concept approaches that leverage near-field enhancements have been realized using precise experimental instrumentation [10–13], but scaling up these approaches remains challenging [14].

Alternative near-field extraction schemes using high refractive index ($n$) waveguides that support large wavevectors relative to the emitter have been proposed [15]. This enables the transmission of free-space evanescent into propagating waves and therefore increases the total radiative transport. Yu, *et. al.* demonstrated this concept by placing a ZnSe hemispherical dome onto a small, flat carbon emitter to achieve an enhancement of ~ 4.5 fold [16]. However, high-index waveguides limit the maximum near-field enhancement because they do not transmit wavevectors that are greater than the light-line in the medium ($k > n\frac{\omega}{c}$).

To overcome this limitation, Messina, *et al.* investigated the use of hyperbolic metamaterial (HMM) waveguides, which have a negative permittivity in either the in-plane ($\epsilon_\parallel < 0, \epsilon_\perp > 0$) or cross plane ($\epsilon_\parallel > 0, \epsilon_\perp < 0$) direction [17]. Due to their unique optical properties, HMM waveguides can support infinitely large wavevectors. The physical transport mechanism entails the coupling of surface plasmon (SPP) and surface phonon polaritons (SPhP) between the emitter and the waveguide [18–20]. In their study, the HMM waveguide was treated as a hypothetical effective medium and was isothermally grounded to the receiver. This simplified model did not consider sources from the waveguide itself, which would arise in a 3-body system when a temperature gradient is established at steady state. The presence of optical loss and gain is widely studied in non-Hermitian optical materials because it can give rise to unique properties such as loss-induced transparency [21]. Furthermore, prior modeling of this system approximated the electromagnetic response of the materials within the waveguide using effective media assumptions. However, it can be expected that the real response of materials and the finite periodicity of the waveguide structure will have significant effects on the heat transfer characteristics.

Here, we investigate the role of sources and real material properties within a non-isothermal HMM that functions as a waveguide between a semi-infinite emitter and receiver. We describe radiative transport in this 3-body system in more detail by taking into account the emissive contributions from the emitter and the HMM waveguide to the receiver across distances of 10-10$^3$ nm. These separation distances represent a transitional regime where ballistic and diffusive transport [22] are both important. By considering sources from the waveguide itself, there arises a balance of optical loss from decreased transmission and optical gain from thermal emission.

In Section II, we revisit the 3-body system proposed by Messina, *et al.* [17] (see Fig. 1a), which consists of a hypothetical HMM placed between two SiC boundaries. Our results show that the negative effects of loss in the HMM waveguide, associated with optical absorption, are compensated by gain from the waveguide sources. Additionally, we show that in some cases, the gains can exceed the losses resulting in an unusual scenario where heat transfer decreases at smaller separations between the emitter and receiver (Section III). Next, we replace the hypothetical HMM waveguide with a more realistic graphene-based HMM (see Fig. 2a). This allows us to investigate near-field thermal extraction in a realizable system [23], where the optical properties of the boundaries are mismatched from the HMM waveguide. Studying the spectral transport in this system reveals that selective near-field transport imposed by the boundary is maintained, even when the emissive contributions from the graphene-based waveguide are the same order of magnitude as the boundary.

Our results suggest that near-field thermal extraction is not as heavily constrained by optical losses or mismatched optical properties. Our final system represents a 3-body system in a multi-layer geometry inspired by tunable plasmonic materials, such as graphene. With our heat transfer model, we expect that one could develop an optimized 3-body system, exhibiting near-field thermal radiation that exceeds contributions from thermal conduction, which is advantageous for several technologies, including thermophotovoltaics (TPVs). In general, the principles developed in our study can be used to understand radiative transport in 3-body systems, which is important for near-field thermal switches [24], transistors [25], sensors [9], and energy converters [26,27].

## II. EMISSIVE CONTRIBUTIONS FROM A NON-ISOTHERMAL HMM WAVEGUIDE

Here, we revisit near-field thermal extraction in a 3-body system that was considered by Messina, *et al* [17]. The system includes a SiC emitter (1) and receiver (3), mediated by a HMM waveguide (2) of thickness $\delta$ (see Fig. 1a). To be consistent with the hypothetical configuration presented in Ref. [17], the HMM is separated from the SiC boundaries by a vacuum gap of thickness $d = 100\ nm$ on either side of the HMM. The optical properties of SiC are modeled using

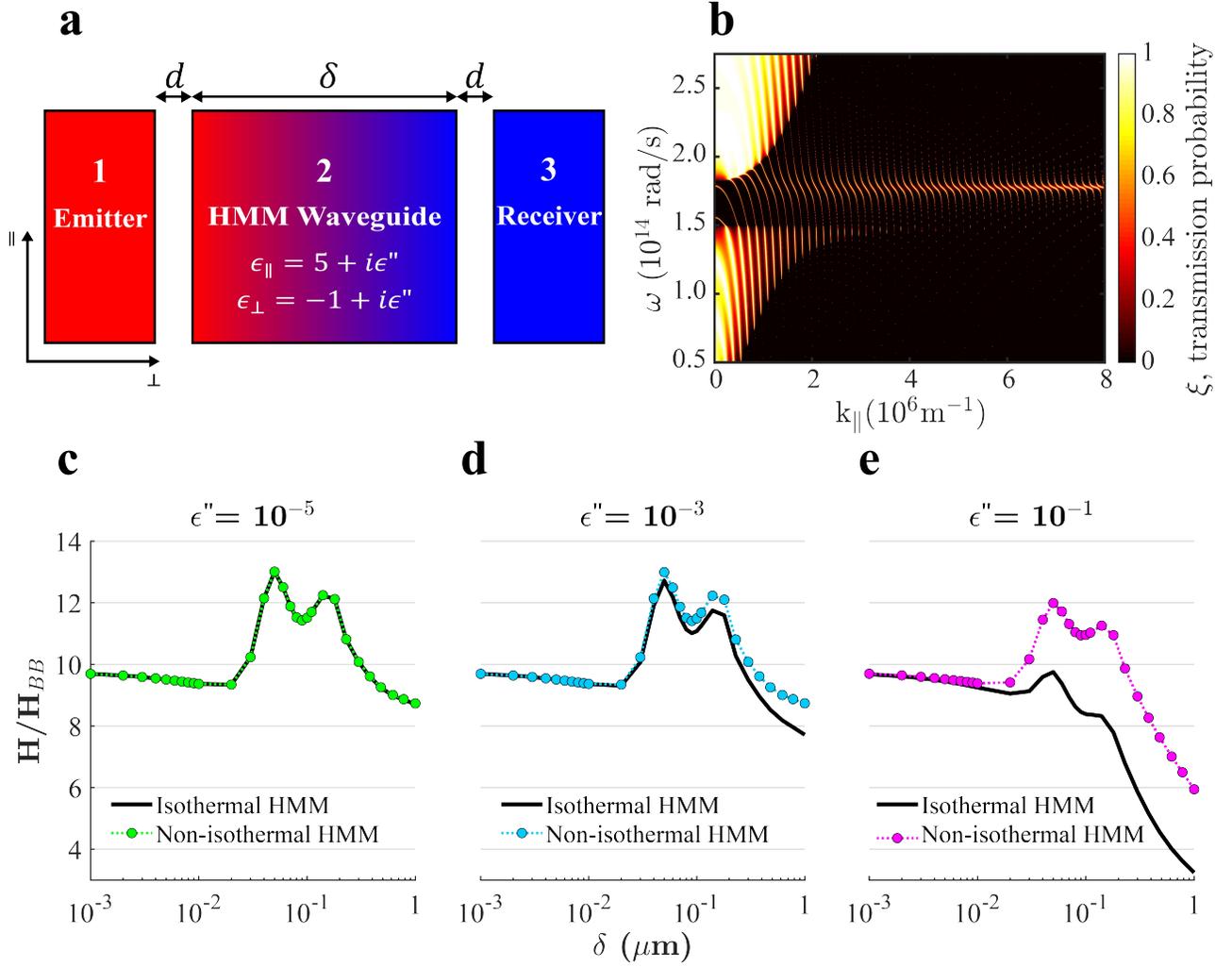

**Figure 1: (a)** Schematic of the 3-body system consisting of a type-I HMM sandwiched between a SiC emitter and receiver and separated by vacuum gaps. **(b)** Energy transmission probability dispersion for the 3-body system. **(c-e)** Near-field enhancement for varying degrees of imaginary permittivity

a Drude-Lorentz approach. The waveguide is modeled as a type-I HMM with the following constant optical properties: $\epsilon_\parallel = 5 + i\epsilon''$ and $\epsilon_\perp = -1 + i\epsilon''$. The imaginary permittivity $\epsilon''$ is treated as a variable parameter to assess the effects of loss from optical absorption.

Heat supplied to the emitter will radiate and conduct along the system, ultimately reaching the receiver. More specifically, radiative heat transfer from the emitter (1) to the receiver (3) can be understood in terms of two contributions: (i) direct radiative transfer $q_{1\to3}$, and (ii) indirect radiative transfer mediated by the HMM waveguide (2), whereby heat conducts/radiates from the emitter to the waveguide, followed by emission from the waveguide to the receiver $q_{2\to3}$. In our model, we approximate the steady state temperature gradient across the HMM as linear. The assumption of a linear temperature gradient is valid for optically thick media or when conduction is the dominant mode of heat transport (see SI Note 2) [28,29]. We also note that the heat transfer from the emitter to the waveguide $q_{1\to 2}$ is approximately equal to $q_{2\to 3}$ because of the symmetry of our material system (see SI Note 3).

The total radiative heat transport in the system is described by,

$$q_{total} = q_{1\to 3} + q_{2\to 3}. \qquad (1)$$

We apply a fluctuational electrodynamics approach to determine the total heat flux from each component to the receiver,

$$q_{1,2\to 3} = \left(\frac{1}{4\pi^2}\right)\int_0^\infty [\theta(\omega,T_{1,2}) - \theta(\omega,T_3)]\Phi_{1,2\to 3}(\omega)\,d\omega, \qquad (2)$$

where $\omega$ is the frequency, $T_{1,2,or\,3}$ is the temperature of each component, $\theta$ is the photon energy distribution, and $\Phi$ is the flux spectrum. The photon energy distribution component is the product of the photon energy $\hbar\omega$ and the Bose-Einstein distribution [30],

$$\theta(\omega,T) = \frac{\hbar\omega}{\exp\left(\frac{\hbar\omega}{k_B T}\right)-1}, \qquad (3)$$

where $\hbar$ is Plank's constant and $k_B$ is Boltzmann's constant.

An analytical expression for the flux spectrum $\Phi$ can be obtained by utilizing the dyadic Green's function to calculate the electromagnetic flux from thermally driven current fluctuations [1,31,32]. In a multilayer geometry, this analytical expression can be solved in conjunction with a scattering matrix method to account for each layer's optical properties, thicknesses, etc. We use the Multilayer Electromagnetic Solver for Heat Transfer (MESH) simulation package [33] to implement this calculation (see SI Note 1) and model the energy transmission probability $\xi$ which is related to the flux spectrum through integration over all in-plane wavevectors $k_\parallel$.

$$\Phi(\omega) = \int_0^\infty \xi(\omega) k_\parallel\, dk_\parallel \qquad (4)$$

The energy transmission probability $\xi$ represents the likelihood of an excited photon being transmitted to the receiver. It can be used to determine mode-specific transmission pathways, such as the excitation of SPPs on graphene or SPhPs on SiC. Using MESH, we modeled the energy transmission probability of the 3-body system (Fig. 1b) and verified that it matched the analytical expression in reference [ref. 17].

In MESH, we discretize the HMM waveguide into 10-nm increments and sum the resulting heat fluxes to the receiver to get a higher resolution heat flux from the waveguide. For consistency, we report our results as a heat transfer coefficient $H$ at 300 K, determined by

$$H = \frac{q_{total}}{T_1 - T_3}. \tag{5}$$

Here, we set the average temperature of the system to 300 K, such that $T_1$ and $T_3$ are equally offset from $T = 300\ K$. Because $T_1 - T_3$ is chosen to be relatively small (20K), we neglect the temperature dependence of the optical properties. To determine a near-field enhancement, we normalize this value by the black-body heat transfer coefficient $H_{BB}$.

With this model, we investigate the role of emissive contributions from the waveguide by comparing the near-field enhancement in two systems: one that has an isothermal HMM waveguide and one that has a non-isothermal HMM waveguide (see Fig. 1c-1e). We vary the magnitude of the imaginary permittivity $\epsilon''$ of the HMM waveguide to investigate how loss can affect the near-field enhancement. First, we reproduce the results of all systems with an isothermal waveguide (black curves in Fig 1c-1e) from ref [17]. In this work, Messina *et al.* describe the peaks in the near-field enhancement around $\delta = 50$ nm and 150 nm as the first and second Fabry-Perot modes that exist due to the finite-nature of the waveguide.

For the system with small losses ($\epsilon'' = 10^{-5}$, Fig. 1c), the near-field enhancement for the isothermal and non-isothermal systems are indistinguishable. In this state, neither losses due to waveguide absorption, nor contributions from thermal emission in the waveguide, have an observable effect on the near-field enhancement. As the imaginary permittivity increases to $\epsilon'' = 10^{-3}$ (Fig. 1d), however, there is a noticeable separation in near-field enhancement between the isothermal and non-isothermal system. For the isothermal system (solid, black line), a higher magnitude of emission from the emitter is absorbed by the waveguide, preventing full transmission to the receiver. When emissive contributions from the non-isothermal HMM waveguide are included (dashed, blue line), the near-field enhancement nearly recovers to its original magnitude (see Fig. 1c). The difference between the two cases is most apparent for the system with an imaginary permittivity of $\epsilon'' = 10^{-1}$ (Fig. 1e). At a waveguide thickness of 1 µm, the emissive contributions from the waveguide account for over 45% of the heat transport to the receiver.

In summary, the emissive contributions from the non-isothermal HMM waveguide compensate for ohmic transmission losses from the emitter to the receiver. This finding suggests that larger near-field enhancements can be extended to longer separation distances, which is important when considering heat transfer mediated by realistic, lossy materials. In this system, the vacuum gaps limit the near-field enhancement (see SI Note 4) compared to the same system with the gaps removed. In the following section, we remove these gaps and replace the ideal HMM with a more realistic graphene-based HMM.

### III. NEAR-FIELD ENHANCMENT IN A REALISTIC 3-BODY SYSTEM

In this section, we investigate the near-field enhancement in a continuous 3-body system to represent a realistic heterogenous structure that does not include nanoscale vacuum gaps. Here, we keep the boundaries as SiC, but replace the hypothetical HMM waveguide with a graphene-based HMM (see Fig. 2a). Graphene has emerged as a candidate material for HMM fabrication because it can be chemically or electronically doped to support SPPs in the mid-infrared [23,34–

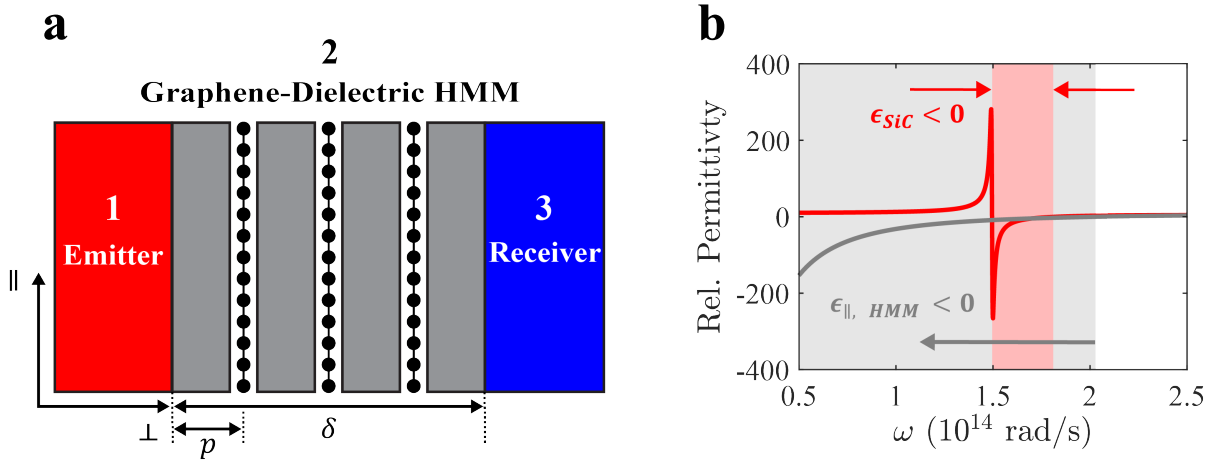

**Figure 2:** (a) Schematic of the continuous 3-body system consisting of a graphene-dielectric HMM sandwiched between a SiC emitter and receiver. (b) The real part of the in-plane permittivity of the HMM (gray, solid line) is plasmonic below angular frequencies of $\sim 2 \cdot 10^{14}$ rad/s (gray area). The real part of the permittivity of the SiC boundaries (red, solid line) is below zero in the overlapping frequency range (red area) due to the presence of a phonon.

36], enabling large thermal fluxes. Additionally, graphene's weak out-of-plane van der Waals (vdW) bonds can lead to high interfacial thermal resistance to phonon-mediated conduction [37,38], which can be beneficial in radiative applications [39]. Our graphene HMM is based on a demonstration of an alternating graphene-dielectric heterostructure with a periodicity of $p = 10\ nm$ [23]. Here, we label graphene as the *active* layer. The optical properties of graphene are taken from a semi-empirical model fit to ellipsometry measurements of chemically doped graphene (see SI Note 5). We treat the dielectric as a weakly absorbing layer that has a constant permittivity of $\epsilon = 10 + 0.01i$. This is representative of many semiconductors, e.g., Si and Ge, in their infrared transmission window. We label the dielectric layers as *passive* because their contributions to near-field emission are small in comparison to the graphene layers.

In general, near-field heat transport in a 3-body system can be optimized if the active layer in the HMM waveguide supports the same surface polaritons as the boundaries. As theoretically demonstrated by Iizuka, *et al.*, the surface polaritons from the boundary can be transported like a relay system from interface to interface of the active layers in the HMM to the receiving boundary. This transport mechanism results in a multibody effect, maximizing near-field transport over the frequency range corresponding to the surface polariton [40]. However, in our system, the boundaries and the waveguide have mismatched optical properties and, therefore, mismatched near-field emission and absorption spectra. This represents a practical implementation of the hypothetical system studied in Section II and Ref. [17].

There is an overlapping frequency range where the real part of the effective, in-plane permittivity of the HMM and the SiC boundaries are both negative due to the excitation of their respective polaritons (see Fig. 2b). This leads to a hybrid polariton coupling mechanism as shown in energy transmission probability across 3 different structures (see Fig. 3a-c):

(a) a SiC emitter and receiver separated by a 100 nm vacuum gap,

(b) a 100 nm thick graphene-dielectric HMM, and

(c) a SiC emitter and receiver separated by a 100 nm thick graphene-dielectric HMM.

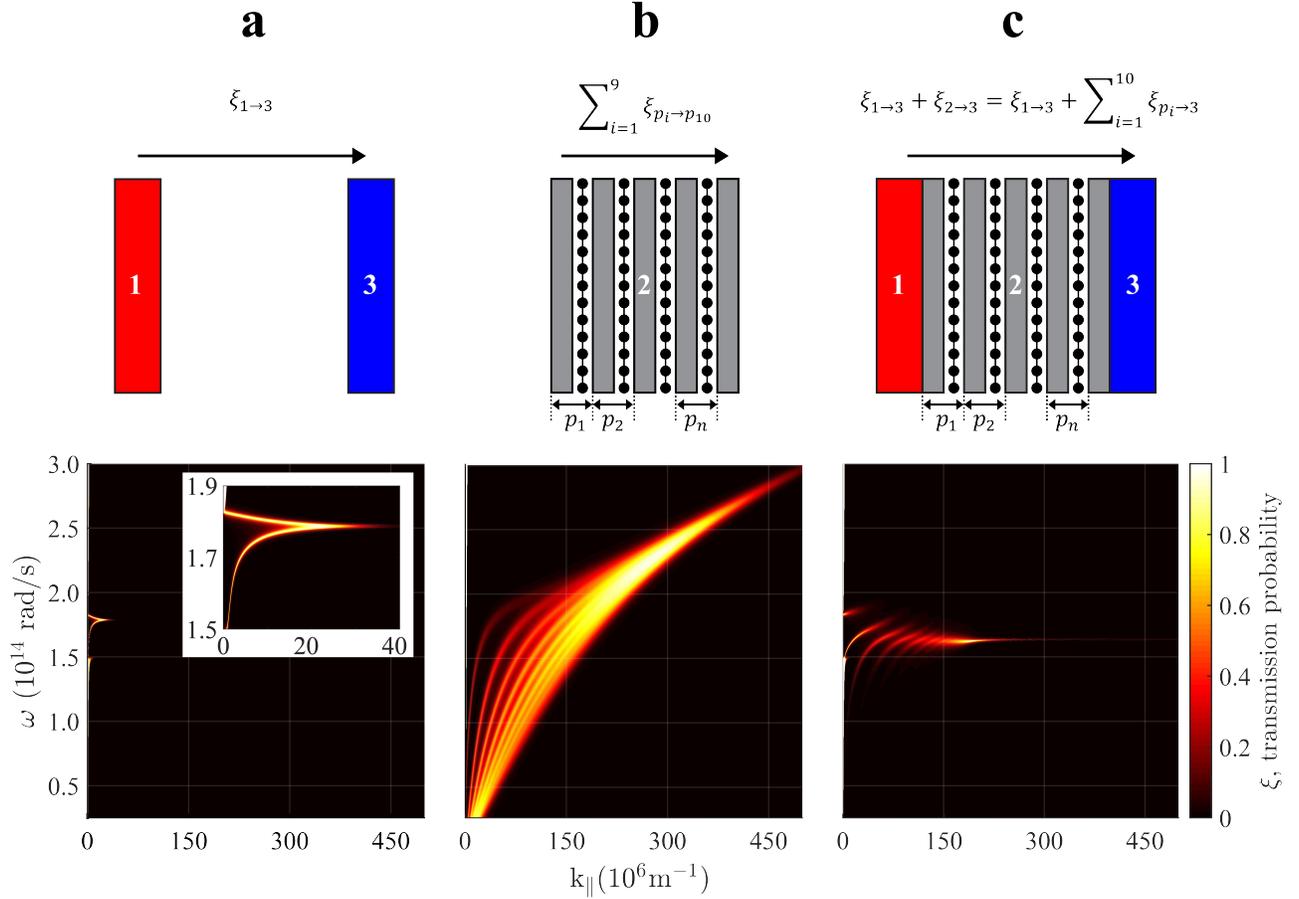

**Figure 3:** Energy transmission probability dispersion diagrams for **(a)** a SiC emitter and receiver separated by a 100 nm vacuum gap, **(b)** a 100 nm thick graphene-dielectric HMM, and **(c)** a SiC emitter and receiver separated by a 100 nm thick graphene-dielectric HMM.

The near-field transport between two SiC boundaries separated by a vacuum gap is dominated by the coupling of two SPhPs, as evidenced by a near-unity band in the transmission probability (see Fig. 3a) over the corresponding phonon frequency range. A much broader, near-unity transmission probability is observed for the graphene-dielectric HMM (see Fig. 3b), extending to very large in-plane wavevectors. The multiband transmission probability, also referred to as the multibody effect, arises from the coupling of numerous symmetric and antisymmetric SPPs between interfaces in the HMM (see SI Note 6) [41,42]. This transmission probability illustrates the broadband nature of HMMs, which has been explored for broadband near-field emission/absorption [43] and near-field thermal extraction [44,45].

In our 3-body system (Fig. 3c), the transmission probability is shaped by both the boundaries and the waveguide. We observe that the transport is selective to the SPhP frequency range from the SiC boundaries. Yet, in this range, the transmission probability resembles the multiband phenomena present due to SPP coupling in the HMM waveguide.

This effect reveals the versatility of having a 3-body system consisting of boundaries and a waveguide with mismatched optical properties. Specifically, the boundaries can be utilized to set the selective frequency range over which transport occurs, while the waveguide can act as a spectrally selective filter to block or transmit the boundaries' surface polaritons based on the hyperbolic band ($\epsilon_{HMM} < 0$) of the HMM waveguide. This mechanism may be useful in near-field TPVs, which suffer from parasitic absorption of low energy, below-bandgap surface phonons [46].

We use the formalism described in Section II to simulate the near-field enhancement at 300K for the 3-body system (see Fig. 4). However, for this system, we do not need to discretize the HMM waveguide because we do not treat it as a bulk, effective medium. Instead, we determine the energy transmission probability from each individual graphene and dielectric layer in the HMM to the receiver. As in Section II, we assume a linear temperature gradient.

Figure 4a shows the near-field enhancement $\left(\frac{H}{H_{BB}}\right)$ as a function of waveguide thickness for the continuous 3-body system consisting of a graphene-dielectric HMM sandwiched between a SiC emitter and receiver. The emissive contributions from the HMM waveguide are larger than the contributions from the SiC boundaries. For HMM thicknesses greater than 100 nm, the emissive contributions from the HMM account for over 90% of the total radiative transport. This is interesting considering that the emission spectrum is characteristic of the SPhP frequency range for SiC.

Furthermore, the heat transfer dependence on separation distance of this system is atypical. The emissive contributions from the HMM waveguide peak at a thickness of $\delta \approx 80$ nm, leading to a non-monotonic trend in the total near-field enhancement. This peak is likely due to the existence of two competing mechanisms. As the waveguide thickness increases, more graphene sheets are introduced into the system, and the number of symmetric and antisymmetric SPP modes contributing to the near-field enhancement increases. This effect is illustrated by the energy transmission probability diagrams in Figure 4b, accompanied by spectral heat transfer plots of the contribution to the near-field enhancement by the HMM waveguide. Simply put, as the dominant source of emission (the HMM) increases in size, so do the contributions coming from it. Further, as the number of graphene sheets in the system increase, the waveguide becomes increasingly lossy, and near-field contributions from HMM periods closer to the emitter are absorbed before reaching the receiver. We note that the waveguide thickness which maximizes $H$ will change with periodicity (see SI Note 7). Furthermore, smaller periods generally lead to a larger near-field enhancement, approaching a threshold at very small $p$.

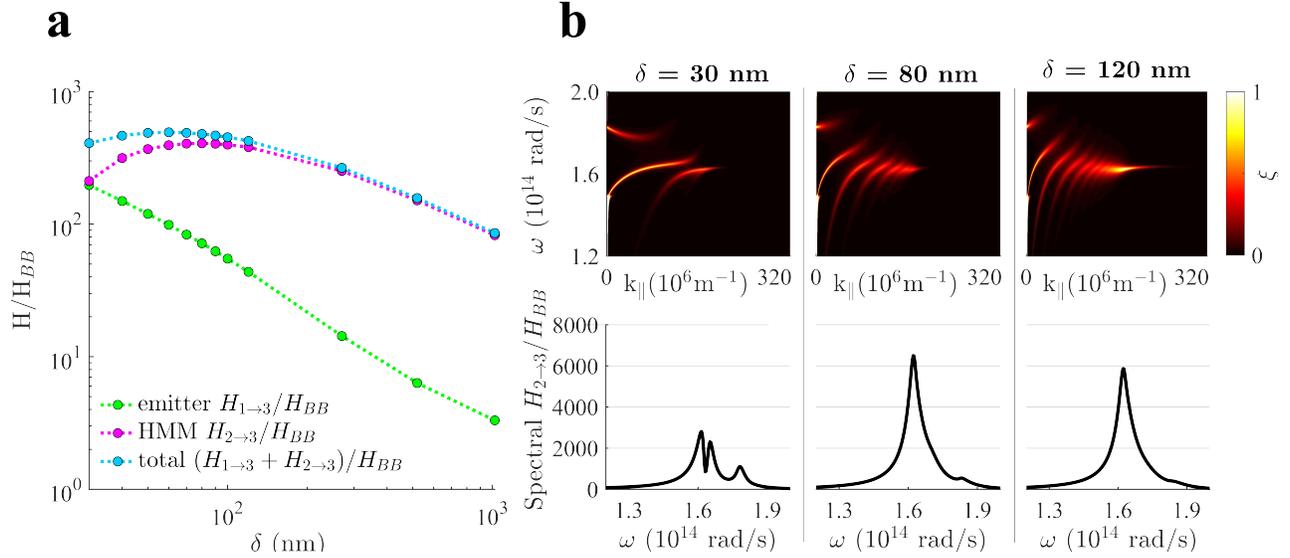

**Figure 4: (a)** Near-field enhancement as a function of waveguide thickness for the continuous 3-body system consisting of a graphene-dielectric HMM sandwiched between a SiC emitter and receiver. **(b)** Dispersion diagrams of the energy transmission probability accompanied by spectral heat transfer plots of the near-field enhancement from the HMM waveguide for waveguide thickness of $\delta = 30\, nm, 80\, nm, $ and $120\, nm$.

The magnitude of the near-field enhancement for this system is much larger than for the system in Section II because the nanoscale vacuum gaps were removed to mimic a realistic continuous structure. Although potentially easier to fabricate, this continuous configuration would come at the cost of competing against phonon-mediated thermal conduction in the system. The near-field thermal radiation in the system can be approximately compared to thermal conduction by calculating an effective radiative conductivity, which is done by multiplying the heat transfer coefficient by the waveguide thickness ($k_{rad} = H\delta$). For this system, the radiative conductivity approaches $k_{rad} = 0.5 \cdot 10^{-3} \frac{W}{mK}$. This value is smaller than the estimated phonon-mediated thermal conduction of $k_{cond.} = 0.125 \frac{W}{mK}$, based on the interfacial thermal resistance of graphene [37,38]. This validates our assumption of a linear temperature gradient. However, recent experimental studies have demonstrated that ultra-insulating vdW heterostructures can have thermal conductivities as low as $7 \cdot 10^{-3} \frac{W}{mK}$ due to the combined effect of high interfacial thermal resistances and mismatches in the phonon density of states between stacked layers [47]. We thus expect that one could develop an optimized 3-body system that allows for higher magnitudes of near-field thermal radiation than thermal conduction.

## IV. CONCLUSION

Our model demonstrates that sources within a HMM waveguide can significantly increase the total near-field enhancement in a 3-body system. For our realistic, continuous 3-body system (SiC-HMM-SiC), over 90% of the total emissive contributions originate from the graphene-based HMM. Additionally, we show that the spectral transport of the system is selective to the frequency range over which the surface polariton from the boundaries (emitter and receiver) exist, even though the HMM waveguide is a major source of thermal emission. These results highlight that the boundaries or the waveguide can be engineered to selectively transmit radiation by changing either the surface polariton at the boundary or changing the spectral range for which the HMM has a hyperbolic dispersion. Our work description of near-field radiative transfer in systems with interposed HMMs could be used to improve spectral selectivity and enhance heat flux.

## ACKNOWLEDGEMENTS

We thank Profs. Theodore Norris, Nicholas Kotov, and Ronald Larson for helpful discussions. This material is based upon work supported by the National Science Foundation (NSF) Graduate Research Fellowship under Grant No. NSF DGE 1256260 and the 3M Foundation Nontenured Faculty Award.

# REFERENCES

[1] S. M. Rytov, *Theory of Electric Fluctuations and Thermal Radiation* (Bedford, 1959).

[2] G. T. Papadakis, S. Buddhiraju, Z. Zhao, B. Zhao, and S. Fan, Broadening near-field emission for performance enhancement in thermophotovoltaics, Nano Lett. **20**, 1654 (2020).

[3] S. McSherry, T. Burger, and A. Lenert, Effects of narrowband transport on near-field and far-field thermophotonic conversion , J. Photonics Energy **9**, 1 (2019).

[4] J. K. Tong, W.-C. Hsu, Y. Huang, S. V. Boriskina, and G. Chen, Thin-film 'Thermal Well' Emitters and Absorbers for High-Efficiency Thermophotovoltaics, Sci. Rep. **5**, 10661 (2015).

[5] W.-C. Hsu, J. K. Tong, B. Liao, Y. Huang, S. V. Boriskina, and G. Chen, Entropic and Near-Field Improvements of Thermoradiative Cells, Sci. Rep. **6**, 34837 (2016).

[6] B. Song, A. Fiorino, E. Meyhofer, and P. Reddy, Near-field radiative thermal transport: From theory to experiment, AIP Adv. **5**, (2015).

[7] E. Tervo, E. Bagherisereshki, and Z. Zhang, Near-field radiative thermoelectric energy converters: a review, Front. Energy **12**, 5 (2018).

[8] S. Molesky and Z. Jacob, Ideal near-field thermophotovoltaic cells, Phys. Rev. B **91**, 205435 (2015).

[9] A. C. Jones and M. B. Raschke, Thermal Infrared Near-Field Spectroscopy, Nano Lett. **12**, 1475 (2012).

[10] A. Fiorino, L. Zhu, D. Thompson, R. Mittapally, P. Reddy, and E. Meyhofer, Nanogap near-field thermophotovoltaics, Nat. Nanotechnol. **13**, 806 (2018).

[11] J. DeSutter, L. Tang, and M. Francoeur, A near-field radiative heat transfer device, Nat. Nanotechnol. **14**, 751 (2019).

[12] A. Fiorino, D. Thompson, L. Zhu, B. Song, P. Reddy, and E. Meyhofer, Giant Enhancement in Radiative Heat Transfer in Sub-30 nm Gaps of Plane Parallel Surfaces Nano Lett. **18**, 3711 (2018).

[13] M. P. Bernardi, D. Milovich, and M. Francoeur, Radiative heat transfer exceeding the blackbody limit between macroscale planar surfaces separated by a nanosize vacuum gap, Nat. Commun. **7**, 12900 (2016).

[14] S. M. Nicaise, C. Lin, M. Azadi, T. Bozorg-Grayeli, P. Adebayo-Ige, D. E. Lilley, Y.

Pfitzer, W. Cha, K. Van Houten, N. A. Melosh, R. T. Howe, J. W. Schwede, and I. Bargatin, Micron-gap spacers with ultrahigh thermal resistance and mechanical robustness for direct energy conversion, Microsystems Nanoeng. **5**, 31 (2019).

[15] J. Li, J. Wuenschell, Y. Jee, P. R. Ohodnicki, and S. Shen, Spectral near-field thermal emission extraction by optical waveguides, Phys. Rev. B **99**, 235414 (2019).

[16] Z. Yu, N. P. Sergeant, T. Skauli, G. Zhang, H. Wang, and S. Fan, Enhancing far-field thermal emission with thermal extraction, Nat. Commun. **4**, 1730 (2013).

[17] R. Messina, P. Ben-Abdallah, B. Guizal, M. Antezza, and S. A. Biehs, Hyperbolic waveguide for long-distance transport of near-field heat flux, Phys. Rev. B **94**, 104301 (2016).

[18] I. Avrutsky, I. Salakhutdinov, J. Elser, and V. Podolskiy, Highly confined optical modes in nanoscale metal-dielectric multilayers, Phys. Rev. B **75**, 241402(R) (2007).

[19] S. V. Zhukovsky, A. Andryieuski, J. E. Sipe, and A. V. Lavrinenko, From surface to volume plasmons in hyperbolic metamaterials: General existence conditions for bulk high-k waves in metal-dielectric and graphene-dielectric multilayers, Phys. Rev. B **90**, 155429 (2014).

[20] S. V. Zhukovsky, O. Kidwai, and J. E. Sipe, Physical nature of volume plasmon polaritons in hyperbolic metamaterials, Opt. Express **21**, 14982 (2013).

[21] H. Jing, Ş. K. Özdemir, Z. Geng, J. Zhang, X.-Y. Lü, B. Peng, L. Yang, and F. Nori, Optomechanically-induced transparency in parity-time-symmetric microresonators, Sci. Rep. **5**, 9663 (2015).

[22] E. E. Narimanov and I. I. Smolyaninov, Beyond Stefan-Boltzmann Law : Thermal Hyper-Conductivity, CLEO Tech. Dig. 2 (2012).

[23] Y.-C. Chang, C.-H. Liu, C.-H. Liu, S. Zhang, S. R. Marder, E. E. Narimanov, Z. Zhong, and T. B. Norris, Realization of mid-infrared graphene hyperbolic metamaterials, Nat. Commun. **7**, 10568 (2016).

[24] D. Thompson, L. Zhu, E. Meyhofer, and P. Reddy, Nanoscale radiative thermal switching via multi-body effects, Nat. Nanotechnol. **15**, 99 (2020).

[25] P. Ben-Abdallah and S.-A. Biehs, Near-Field Thermal Transistor, Phys. Rev. Lett. **112**, 044301 (2014).

[26] K. Chen, T. P. Xiao, P. Santhanam, E. Yablonovitch, and S. Fan, High-performance near-

field electroluminescent refrigeration device consisting of a GaAs light emitting diode and a Si photovoltaic cell, J. Appl. Phys. **122**, (2017).

[27] L. Zhu, A. Fiorino, D. Thompson, R. Mittapally, E. Meyhofer, and P. Reddy, Near-field photonic cooling through control of the chemical potential of photons, Nature **566**, 239 (2019).

[28] M.-J. He, H. Qi, Y.-F. Wang, Y.-T. Ren, W.-H. Cai, and L.-M. Ruan, Near-field radiative heat transfer in multilayered graphene system considering equilibrium temperature distribution, Opt. Express **27**, A953 (2019).

[29] M. F. Modest, in *Radiat. Heat Transf.*, Third Edit (Elsevier Inc, 2013), pp. 454–479.

[30] P. Wurfel, The chemical potential of luminescent radiation, J. Phys. C Solid State Phys. **15**, 3967 (1982).

[31] D. Polder and M. Van Hove, Theory of Radiative Heat Transfer between Closely Spaced Bodies, Phys. Rev. B **4**, 3303 (1971).

[32] S. M. Rytov, Y. A. Kravtsoy, and V. I. Tatarskii, *Priniciples of Statistical Radiophysics (Vol 3) Elements of Random Fields* (Springer, Berlin, Heidelberg, 1989).

[33] K. Chen, B. Zhao, and S. Fan, MESH: A free electromagnetic solver for far-field and near-field radiative heat transfer for layered periodic structures, Comput. Phys. Commun. **231**, 163 (2018).

[34] M. A. K. Othman, C. Guclu, and F. Capolino, Graphene-based tunable hyperbolic metamaterials and enhanced near-field absorption, Opt. Express **21**, 7614 (2013).

[35] I. V. Iorsh, I. S. Mukhin, I. V. Shadrivov, P. A. Belov, and Y. S. Kivshar, Hyperbolic metamaterials based on multilayer graphene structures, Phys. Rev. B **87**, 075416 (2013).

[36] S. Dai, Q. Ma, M. K. Liu, T. Andersen, Z. Fei, M. D. Goldflam, M. Wagner, K. Watanabe, T. Taniguchi, M. Thiemens, F. Keilmann, G. C. A. M. Janssen, S.-E. Zhu, P. Jarillo-Herrero, M. M. Fogler, and D. N. Basov, Graphene on hexagonal boron nitride as a tunable hyperbolic metamaterial, Nat. Nanotechnol. **10**, 682 (2015).

[37] E. Pop, V. Varshney, and A. K. Roy, Thermal properties of graphene: Fundamentals and applications, MRS Bull. **37**, 1273 (2012).

[38] D. Estrada, Z. Li, G. Choi, S. N. Dunham, A. Serov, J. Lee, Y. Meng, F. Lian, N. C. Wang, A. Perez, R. T. Haasch, J.-M. Zuo, W. P. King, J. A. Rogers, D. G. Cahill, and E. Pop, Thermal transport in layer-by-layer assembled polycrystalline graphene films, Npj

2D Mater. Appl. **3**, 10 (2019).

[39] D.-H. Lien, S. Z. Uddin, M. Yeh, M. Amani, H. Kim, J. W. Ager, E. Yablonovitch, and A. Javey, Electrical suppression of all nonradiative recombination pathways in monolayer semiconductors, Science (80-. ). **364**, 468 (2019).

[40] H. Iizuka and S. Fan, Significant Enhancement of Near-Field Electromagnetic Heat Transfer in a Multilayer Structure through Multiple Surface-States Coupling, Phys. Rev. Lett. **120**, 063901 (2018).

[41] B. Wang, X. Zhang, X. Yuan, and J. Teng, Optical coupling of surface plasmons between graphene sheets, Appl. Phys. Lett. **100**, 131111 (2012).

[42] X. L. Liu and Z. M. Zhang, Graphene-assisted near-field radiative heat transfer between corrugated polar materials, Appl. Phys. Lett. **107**, (2015).

[43] Y. Guo, C. L. Cortes, S. Molesky, and Z. Jacob, Broadband super-Planckian thermal emission from hyperbolic metamaterials, Appl. Phys. Lett. **101**, 131106 (2012).

[44] J. Shi, B. Liu, P. Li, L. Y. Ng, and S. Shen, Near-Field Energy Extraction with Hyperbolic Metamaterials, Nano Lett. **15**, 1217 (2015).

[45] B. Liu and S. Shen, Broadband near-field radiative thermal emitter/absorber based on hyperbolic metamaterials: Direct numerical simulation by the Wiener chaos expansion method, Phys. Rev. B **87**, 115403 (2013).

[46] K. Chen, P. Santhanam, and S. Fan, Suppressing sub-bandgap phonon-polariton heat transfer in near-field thermophotovoltaic devices for waste heat recovery, Appl. Phys. Lett. **107**, 091106 (2015).

[47] S. Vaziri, E. Yalon, M. Muñoz Rojo, S. V Suryavanshi, H. Zhang, C. J. McClellan, C. S. Bailey, K. K. H. Smithe, A. J. Gabourie, V. Chen, S. Deshmukh, L. Bendersky, A. V Davydov, and E. Pop, Ultrahigh thermal isolation across heterogeneously layered two-dimensional materials, Sci. Adv. **5**, eaax1325 (2019).

[48] See Supplemental Material at [insert URL]